\documentclass[11pt]{article}  
\usepackage{cancel} 
\usepackage{simplewick} 
\usepackage{fullpage} 	
\usepackage{amssymb}     
\usepackage{amsmath}
\usepackage{hyperref} 
\hypersetup{    colorlinks,     citecolor=black,     filecolor=black,    linkcolor=black,     urlcolor=black}  
\usepackage{graphicx,subcaption} 
\usepackage{multirow} 
\usepackage{dsfont} 
\usepackage{cite}       
\usepackage{axodraw4j}
\usepackage{epstopdf}
\usepackage{slashbox,pict2e} 

\newcommand\+{\dagger}

\renewcommand\d{\partial}
\newcommand\<{\langle}
\renewcommand\>{\rangle}
\renewcommand\O{{\mathcal O}}

\newcommand\x{{\bf x}}

\renewcommand\P{{\bf P}}

\begin{document}

\title{\begin{flushright}\vspace{-1in}
       \mbox{\normalsize  EFI-14-27}
       \end{flushright}
       \vskip 20pt
 Operator Product Expansion and Conservation Laws
 in Non-Relativistic Conformal Field Theories}

\date{\today}

\author{
Siavash Golkar\thanks{\href{mailto:golkar@uchicago.edu} 
    {golkar@uchicago.edu}} ~ \hspace{-4pt} and 
   Dam T. Son\thanks{\href{mailto:dtson@uchicago.edu}
     {dtson@uchicago.edu}} \\ \\
   {\it \it Kadanoff Center for Theoretical Physics and Enrico Fermi Institute,}\\
   {\it   University of Chicago, 5640 South Ellis Ave., Chicago, IL 60637 USA}
} 

\maketitle


\begin{abstract}
We explore the consequences of conformal symmetry for the operator
product expansions in nonrelativistic field theories.  Similar to the relativistic case, the
OPE coefficients of descendants are related to that of the primary. However, unlike relativistic CFTs the 3-point function of primaries is not completely specified by conformal symmetry. 
Here, we show that the 3-point function between operators with nonzero
particle number, where (at least) one operator has the lowest
dimension allowed by unitarity, is determined up to a numerical
coefficient. We also look at the structure of the family tree of primaries with zero particle number  and discuss the presence of conservation laws in this sector.
\end{abstract}

\newpage

\section{Introduction}

Experimental studies of fermions at unitarity have stimulated
theoretical developments of nonrelativistic conformal field theories.
The conformal extension of the Galilean algebra was found a long
time ago \cite{Hagen:72,Niederer:72} and was later analyzed in the context of string theory.  Mehen, Stewart, and Wise
explored the consequences of the conformal invariance for the
scattering amplitudes involving unitarity fermions\cite{Mehen:99}, and subsequently other applications have been considered in the literature\cite{Henkel:93,Henkel:03,Son:07}.

The understanding of the operator structure of nonrelativistic
conformal field theory is still not complete.  Only recently has the
operator product expansion begun to be explored.  The motivation was
to probe unitarity fermions at short distances.  There have also been
attempts to construct holographic duals of the unitarity fermions\cite{Son:08,McGreevy:08,Moroz:09}.  A
very interesting nontrivial check is the computation of the
three-point function from hologrpahy, which yields the same result as
the calculation in the field theory of unitarity fermions.

In this paper we explore the consequences of conformal invariance on
the structure of the OPEs.  One property of nonrelativistic OPEs is
that the OPE coefficients involves, in general, nontrivial functions
of the ratio $\x^2/t$, where $\x$ and $t$ are the separation between
the two points:
\begin{equation}
  O_1(x) O_2(0) = \sum
  \frac1{|\x|^{\Delta_1+\Delta_2-\Delta_n}} \, c_n\!\left(\tfrac{x^2}t\right) O_n(0)
\end{equation}
Here and later, $x\equiv(t,\x)$, but $x^2\equiv|\x|^2$.

In general conformal symmetry is not powerful enough to restrict the
form of the functions $c_n$. In this work, however, we show that when
one of the operators participating in the OPE is an elementary field,
that is it carries scaling dimension equal to $d/2$,
the OPE coefficents are, in general, determined up to a numerical
coefficient.  The exception is when one of the other operator has
particle number zero, in which case not all the OPE coefficients are
fixed by symmetry.

We also look at the structure of the family generated by a primary with particle number zero. This case is particular because the a subset of the ladder operators in the algebra commute when the particle number is zero. We discuss the consequences of this degeneracy and its implications for the existence of conservation laws.

\section{Nonrelativistic conformal symmetry}

To make the paper self-contained, we recall some basic fact about
nonrelativistic conformal field theories\cite{Son:07}.

\subsection{The Schr\"odinger algebra}

The Schr\"odinger algebra in $d$ dimensions, $\mathfrak{schr}_d$ is
formed from the operators $N,$ $D$, $M_{ij}$, $K_i$, $P_i$, $C$ and
$H$, respectively the number (mass), scaling, rotation, Galilean
boost, spatial translation, special Schr\"odinger transformation and
time translation operators. The operator $N$ is central and all
operators transform with the appropriate tensor structure under
rotations $M_{ij}$. The rest of the algebra is given in Table
\ref{tab:commutators}.
\begin{table}[ht]
\centering
\newcommand{\mc}[3]{\multicolumn{#1}{#2}{#3}}
\begin{tabular}{|l|l|l|l|l|l|}\hline
\mc{1}{|c|}{\backslashbox{A}{B}} & \mc{1}{|c|}{$P_j$} & \mc{1}{|c|}{$K_j$} & \mc{1}{|c|}{$D$} & \mc{1}{|c|}{$C$} & \mc{1}{|c|}{$H$} \\ \hline
\mc{1}{|c|}{$P_i$} & \mc{1}{|c|}{$0$} & \mc{1}{|c|}{$-i\delta_{ij}N$} & \mc{1}{|c|}{$-iP_i$} & \mc{1}{|c|}{$-iK_i$} & \mc{1}{|c|}{$0$} \\ \hline
\mc{1}{|c|}{$K_i$} & \mc{1}{|c|}{$i\delta_{ij}N$} & \mc{1}{|c|}{$0$} & \mc{1}{|c|}{$iK_i$} & \mc{1}{|c|}{$0$} & \mc{1}{|c|}{$iP_i$} \\ \hline
\mc{1}{|c|}{$D$} & \mc{1}{|c|}{$iP_j$} & \mc{1}{|c|}{$-iK_j$} & \mc{1}{|c|}{$0$} & \mc{1}{|c|}{$-2iC$} & \mc{1}{|c|}{$2iH$} \\ \hline
\mc{1}{|c|}{$C$} & \mc{1}{|c|}{$iK_j$} & \mc{1}{|c|}{$0$} & \mc{1}{|c|}{$2iC$} & \mc{1}{|c|}{$0$} & \mc{1}{|c|}{$iD$} \\ \hline
\mc{1}{|c|}{$H$} & \mc{1}{|c|}{$0$} & \mc{1}{|c|}{$-iP_j$} & \mc{1}{|c|}{$-2iH$} & \mc{1}{|c|}{$-iD$} & \mc{1}{|c|}{$0$}\\ \hline
\end{tabular}
\caption{Values of $[A,B]$}
\label{tab:commutators}
\end{table}

We look at representations made of local operators such that:
\begin{equation}
 	\O(x)=e^{iHt-i\P\cdot \x} \O(0)
   e^{-iHt+i\P\cdot \x}
\end{equation}
Since $N$ is central, it is also convenient to look at operators that
have specific particle number $N_\mathcal O$:
\begin{equation}
 	[N,\, \O(0)]=N_\O \O(0)
\end{equation}

If we look at operators $C$, $H$ and $D$, they span a subalgebra of
$\mathfrak{schr}_d$ isomorphic to $\mathfrak{su}(1,1)$ corresponding to
the lowering, raising and diagonal operators repectively. An operator
$\mathcal O$ is said to have scaling dimension $\Delta_\mathcal O$ if:
\begin{equation}
 	[D,\, \O(0)]=i\Delta_\O \O(0).
\end{equation}
The operators $P_i$ and $K_i$ also act as raising and lowering for the
eigenvalue of $D$, albeit with increments of 1 (compared to increments
of 2 in the case of $H$ and $C$). We can therefore, classify the
representations of $\mathfrak{schr}_d$ with their number and the
lowest scaling dimension. (These form standard cyclic
representations.) We will call the operator of lowest weight a primary
operator:
\begin{equation}
 	[C,\,\mathcal O]=0,\;[K_i,\,\O]=0
\end{equation}
For primary operators, the action of $C$ and $K_i$ can be written as:
\begin{align}
  [K_i,\,\mathcal O(x)]&=(-it\partial_i+N_\mathcal O x_i)\mathcal O,\notag\\
  [C,\,\mathcal O(x)]&=-i(t^2\partial_t+x_i\partial_i+t\Delta_\mathcal O)
  \mathcal O+\frac {x^2}2N_\mathcal O \mathcal O.
\end{align}

\subsection{Correlators}
The Schr\"odinger algebra puts restrictions on the form of the
correlators. Similar to relativistic CFTs the form of the 2-point
function of primaries is determined upto an overall constant\cite{Henkel:93}. For
scalar primaries we have (similar result holds for operators which
transform like tensors with respect to rotations):
\begin{equation}
\label{eq:2pt}
  \< \mathcal O_1(t_1,\x_1) \O_2(t_2,\x_2) \> = 
  c \,\delta_{\Delta_1\Delta_2}t_{12}^{-\Delta_1}
  \exp\left(\frac{iM}2 \frac{\x_{12}^2}{t_{12}}\right).
\end{equation}
where $t_{12}\equiv t_1-t_2$ and $\x_{12}\equiv\x_1-\x_2$.

However, conformal symmetry does not completely fix the three-point
function, which can depend on one arbitrary function (in the
relativistic case, this occurs for four- and higher-point correlators).  We have,
for three- and four-point functions \cite{Henkel:93,Volovich:09}:
\begin{align}
  \langle \O_1 \O_2 \O_3\> &= 
   \prod_{i<j} t_{ij}^{\frac12\Delta-(\Delta_i+\Delta_j)} 
   \exp\left(\frac{iM_1}2 \frac{x_{13}^2}{t_{13}}+\frac{iM_2}2 
   \frac{x_{23}^2}{t_{23}}\right) F_3(v_{123})\notag\\
  \< \O_1 \O_2 \O_3 \O_4 \> &= 
    \prod_{i<j} t_{ij}^{\frac16\Delta-(\Delta_i+\Delta_j)/2} 
    \exp\left(\frac{iM_1}2 \frac{x_{14}^2}{t_{14}}+\frac{iM_2}2 
    \frac{x_{24}^2}{t_{24}}+\frac{iM_2}2 \frac{x_{34}^2}{t_{34}}\right) 
       \times\notag\\
  &\qquad\times F_4\left( \frac{t_{12}t_{34}}{t_{13}t_{24}},
    v_{124},v_{134},v_{234}\right)
\label{eq:3-4pt}
\end{align}
where $\Delta=\sum \Delta_i$ and:
\begin{equation}
 v_{ijk}=\frac12\left(\frac{x^2_{jk}}{t_{jk}}- \frac{x^2_{ik}}{t_{ik}}
  + \frac{x^2_{ij}}{t_{ij}}\right).
\end{equation} 
$F_3$ and $F_4$ are functions of one and four variables
respectively. The form of these functions is not not restricted by the symmetries.

\section{The operator product expansion}

Similar to relativistic CFTs \cite{Zimmermann:70,Wilson:71,Wilson:72}, we expect the product of two operators
at two different points to be expressible as a sum of local
operators. Restricting to the OPEs of primary scalar operators
(similar results hold for the OPE of tensor operators) we have:
\begin{equation}
 	\mathcal O_i(t,x)\mathcal O_j(0)
      = \sum_{k,l} C_{ij}^{kl}(t,x) \mathcal O_{k,l}\,,
	\label{eq:OPE-exp}
\end{equation}
where $\mathcal O_{k,l}$ denotes the $l$'th descendant of the primary
operator $\mathcal O_k$.  The Sch\"odinger algebra puts stringent
restrictions on the form of the coefficients $C_{ij}^{kl}$, where in
most cases, we can read off the coefficient of descendants from the
coefficient of the primary $C_{ij}^{k0}$. We assume the operator $\mathcal O_k$ has nonzero operator number. The case of operators carrying zero particle number will be discussed in section \ref{sec:NZO}.

 Here, we recall the procedure for deriving these coefficients. The first few terms of the expansion \eqref{eq:OPE-exp} can be written as:
\begin{equation}
  \mathcal O_2(x)\mathcal O_3(0) 
  = (C_0(x) + C_1^i(x) \partial_i + C_2(x) \partial_t 
  + C_3^{ij}(x) \partial_i \partial_j + \cdots)\mathcal O_1.
\label{eq:OPE-fft}
\end{equation}
Commuting both sides with $K_i$, we get:
\begin{align}
\label{eq:C-relations}
   (-it\d_i + N_3 x_i) C_0 &= N_1 C_1^i,\notag\\
   (-it\d_i + N_3 x_i) C_1^j &= -i\delta_{ij}C_2 + 2N_1 C_3^{ij},
\end{align}
while commuting with $C$ gives:
\begin{equation}\label{eq2}
  \left( -it^2\d_t -itx_i\d_i -it\Delta_3 + \frac{x^2}2N_3\right)C_0
  = -i\Delta_1C_2 +N_1 C_3^{ii}.
\end{equation}
These equations completley determine $C_i^i$, $C_2$, and $C_3^{ij}$ in
terms of $C_0$:
\begin{align}
\label{eq:C_1,2}
  C_1^i =& \frac1{N_1} (-it\d_i + N_3 x_i) C_0\notag\\
  C_2 =& \frac i{N_1(2\Delta_1-d)} \left[ t^2(-2iN_1\d_t+\d^2)
  - 2i N_2tx_i\d_i + i(N_3d-2N_1\Delta_3) t 
  + N_3 N_2 x^2\right] C_0
\end{align}
Note that $N_2+N_3=N_1$.  The expression for $C^{ij}_3$ in terms of
$C_0$ can be easily written down using \eqref{eq:C-relations} and \eqref{eq:C_1,2}. The rest of the coefficients in the series can be derived in a similar fashion.

\subsection*{Operators with dimension $d/2$}

From \eqref{eq:C_1,2} it is clear that when the dimension of
the operator $\mathcal O_1$ is
equal to $d/2$, the equation  for $C_2$ becomes
ill-defined and can be interpreted as a restriction on
$C_0$ itself.  We note that the value $d/2$ is the unitarity bound on operator dimensions.
In the theory of fermions at unitarity, the elementary fermion field
$\psi$ and its Hermitian conjugate $\psi^\dagger$ have this dimension.  In this case, we can identify $\O_1=\psi$, so $N_1=-1$, $\Delta_1=d/2$, and
$N_3=-(N_2+1)$.  The equation for $C_0$ is then:
\begin{equation}
    \left[ t^2(2i\d_t+\d^2)
  - 2i N_2 t x_i  \d_i + i(N_3d+2\Delta_3) t 
   +N_3 N_2 x^2\right] C_0=0.
	\label{eq:C_0-DE}
\end{equation}
This is a partial differential equation for $C_0$. However, scale
invariance tells us that $C_0$ has the form:
\begin{equation}
 	C_0(t,x)=t^{-\frac12\Delta_{23,1}} f(x^2/t).
	\label{eq:C_0}
\end{equation}
Using this form, the PDE for $C_0$ turns into an ordinary differential
equation for $f$:
\begin{equation}
  4yf'' + 2 [d-i(N_2-N_3)y] f' + \left[ N_2N_3 y
  -i  \frac d2 (N_2-N_3) - i(\Delta_2-\Delta_3)\right] f = 0.
\end{equation}
The solution to this equation is
\begin{equation}
f_0= e^{-\frac{i}{2} N_3 y} \left[A \, U(-\alpha,\frac d2,-\frac{i y}{2})
 + B \, L_{\alpha}^{\frac d2 -1}(-\frac{i y}{2})\right],
\end{equation}
where $\alpha=\frac12 (\Delta_2-\Delta_3-\frac d2)$, $U(a,b,x)$ is the 
confluent hypergeometric funtion, $L_a^b(x)$ is the generalized
Laguerre polynomial and $A$ and $B$ are coefficients determined by
the boundary conditions. Here, we note that regularity at $y\to0$
sets $A$ equal to zero and $B$ can be fixed with proper normalization
requirements. 

The final result is that for the OPE of
any two primary operators, the exact form of the coefficient of
dimension $d/2$ operators is known. In what follows, we discuss the
consequences of this extra information.

It is worth noting that the fact that in equation~\eqref{eq:C_1,2}, instead of deriving the coefficients of the descendants
we ended up with a constraint on the coefficient of the primary is a
reflection of the existence of a \emph{null} operator (an operator which is
both a primary and a descendant). Therefore, if we can find more null
operators we would be able to impose more constraint equations for the OPE
coefficients. With a bit of algebra, one can show that all null operators with non-zero particle number are of the
form $(\partial^2-2iN\partial_t)^n \mathcal O$ where $\mathcal O$ is a
primary of dimension $d/2-n+1$ (This was first shown in \cite{Dobrev:97}. For a recent review see \cite{Dobrev:13}). In the case above $n=1$. Noting that
$D=d/2$ is the lowest scaling dimensions from unitarity constraints,
we see that there is a unique null operator in any Schr\"odinger
symmetric theory (more precisely, the number would be equal to the
number of operators with scaling dimension $d/2$). Again, the case of operators with zero particle number is unique and we examine it in section \ref{sec:NZO}.

\section{Restrictions on multipoint correlators}

\subsection*{3-pt Functions}
One can use the known OPE coefficient to restrict the form of the 3pt
functions which include operators of dimension $d/2$. From equation \eqref{eq:3-4pt} we
have:
\begin{equation}
 	\lim_{x,t\to 0} \langle \O_1(y) \O_2(x)\O_3(0)\rangle=(t_y)^{-\Delta_1}t^{-\frac12\Delta_{23,1}}
	\exp\left[\frac{iM_1}2 \frac{y^2}{t_y}+\frac{iM_2}2 
   \frac{x^2}{t}\right] F_3(x^2/t),
\end{equation}
where the fraction $x^2/t$ is kept finite in the limit. On the other hand from the OPE \eqref{eq:OPE-fft} we have:
\begin{equation}
 	\lim_{x\to 0} \langle \O_1(y) \O_2(x)\O_3(0)\rangle=t^{-\Delta_{23,1}/2} f(x^2/t) (t_y)^{-\Delta_1}\exp\left[i \frac{M_1}{2}\frac{y^2}{t_y}\right],
\end{equation}
where $f$ is defined in equation \eqref{eq:C_0}. From these we conclude that:
\begin{equation}
 	F_3(z)=e^{-\frac i2 M_1 z} f(z).
\end{equation}
Therefore, if one of the operators has dimension $d/2$, the form of the 3pt 
function is completely known. We will identify $\O_1=\psi$, so $N_1=-1$,
$\Delta_1=d/2$, and $N_3=1-N_2$.. There is
one caveat here. The limits must all exist and be non-zero.


Now that we have the 3pt function exactly, it is easy to see that we
can derive any of the OPEs involved by taking the appropriate
limit. In particular, we can write:
\begin{equation}
 	\psi(y) \O_2(x)=D_0(y-x) \O_3^\+(x).
\end{equation}
Then, using the known result of the 3 point function discussed above, we
can relate the coefficient $D_0$ to the coefficient $C_0$ in equation
\eqref{eq:OPE-fft} by taking different limits. We have:
\begin{equation}
 	C_0(x,t)=t^{\Delta_1-\Delta_3} \exp\left[i M_2 \frac{x^2}{2 t}\right] D_0(x,t)
\end{equation}
This can now be used to derive a differential equation for $D_0$ using
\eqref{eq:C_0-DE}:
\begin{equation}
 	\left[2i\d_t+\d^2\right] D_0(x,t)=0,
\end{equation}
where we have used the scaling property of $D_0(x,t)$. We see that the OPE coefficient
satisfies the Schr\"odinger equation. Moreover, we can also show that:
\begin{equation}
 	\left[2i M_1\d_{t_y}+\d_y^2\right] \langle \O_1(y) \O_2(x)\O_3(0)\rangle=0,
\end{equation}
whenever $\Delta_1=d/2$ and $t_y>t_x>0$. Since the 3-point function of primaries was already constrained to a function of a single variable, this extra constraint completely determines the function up to boundary conditions. In fact this explains why the 3-point functions of the elementary particle $\psi$, calculated in different theories\cite{Henkel:03,Moroz:09} match because they all satisfy this differential equation. 

We could have expected this from the OPE expansion of the correlator. We note that the 
two point function of primary operators of scaling dimension $d/2$ is the Green's function 
of the Schr\"oedinger operator. Therefore, any n-point function that can be collapsed down to
the two point function of these operators, will satisfy the Schr\"oedinger equation. This is
possible only when the operator with the scaling dimension $d/2$ appears as either the first
or last operator in a time-ordered correlator.

What happens when the operator appears in the middle of a correlator is not clear in general.
However, in the special case of fermions at unitarity we can easily answer the question. The
classical equation of motion states that:
\begin{equation}
 	\left[\d_t+\frac1{2i M_1}\d^2\right] \psi_\uparrow=\psi_\downarrow^\dagger\phi.
\end{equation}
A simple Schwinger-Dyson type argument shows this result should hold inside correlators
(upto contact terms). This is indeed true and in the case of 3-point functions can be checked 
explicitly using the differential equations derived above. We note that since the operator
$\psi_\downarrow^\dagger\phi$ includes both annihilation and creation operators, its correlators
would vanish if it appears at the endpoints, thus recovering the previous result. 

We expect a similar relation to hold in general where the Schr\"oedinger operator acting on 
a primary operator of scalind dimension $d/2$ would equal a primary operator that includes
both creation and annihilation operators.

\subsection*{4pt Functions}
Knowing the exact form of the 3pt functions, we can now try to
restrict the form of the 4pt function. In particular, it seems that we
are in a similar situation to relativistic CFTs, where the 3pt
function is known and the 4pt can be viewed as a sum over intermediate
states. The easiest way to calculate the form of the 4pt function when
we have two operators of dimension $d/2$ is to calculate it within a
particular theory. Then by the previous results, the answer should be
the same in any other theory with the same 3pt functions.

Here, we will use the embedding of $\mathfrak{schr}_d$ in
$\mathfrak{conf}_{d+2}$ to carry out the calculation. This is a
particularly nice choice, as the structure of 4pt function of the
relativistic CFTs is known and the mapping between the 3pt functions
is simple. Doing the calculation, we obtain:
\begin{multline}
 \< \O_1 \O_2 \O_3\O_4\> \sim \delta(\sum M_i) 
   \int \text d\zeta_1 \text d\zeta_2 \text d\zeta_3 
   e^{-iM_1\zeta_1 -M_2\zeta_2-M_3\zeta_3} \times \\
   \times F\left(\frac{t_{12}t_{34}}{t_{13}t_{24}}  
  \frac{\zeta_3(\zeta_1 - \zeta_2 +iv_{124})}
  {\zeta_2(\zeta_1 - \zeta_3 +iv_{134})}
    , \frac{t_{12}t_{34}}{t_{23}t_{14}}  
  \frac{\zeta_3(\zeta_1 - \zeta_2 +iv_{124})}
  {\zeta_1(\zeta_2 - \zeta_3 +iv_{234})}\right).
\end{multline}
We see that the general function $F_4$ of 4 variables is expressible
as a function of only 2 variables. However, there is no simple
relation between the two and one must use the integral formula above
to relate them.

\section{Zero Particle Number Sector And Conservation Laws} \label{sec:NZO}
In this section we look at the structure of the family associated with a primary $O$ with a zero particle number. What is special in this sector is the fact that the operators $K_i$ and $P_j$ that act as ladder operators, commute with each other. What this means is that descendants that we derive from raising the primary with the $P$ operator cannot be lowered back to the primary by the use of the lowering operator $K$:
\begin{equation}
	K_i P_j O= P_j K_i O + [K_i, P_j] O=0.
\end{equation}
hence, if we have an operator $O'_i$ which lowers to the primary operator $O$ with the use of $K_i$, this operator cannot be a descendant of $O$. We will call these operators which are not descendants of the primaries but nevertheless play a role in the operator algebra, \emph{alien} operators.

We can derive a simple relation between the the family members (alien or descendant) at any level. Assuming $O^{(n)}$ is an alien operator at level n, which lowers to $O_i^{(n-1)}$, we have:
\begin{equation}
	P_j O_i^{(n-1)} = P_j K_i O^{(n)}= K_i P_j O^{(n)}.
\end{equation}
This is just the statement of the commutation of the ladder operators. Normally, the cartoon of a family tree would depict the ladder operators as parallel but in opposite directions. That is the operators $PK$ and $KP$, raising followed by lowering as well as lowering followed by raising,  would act as identity. However, in the zero particle number case, $K_iP_j=P_jK_i$ and it is \emph{not} proportional to the identity. If we orient the tree such that the scaling dimension changes vertically, $K_iP_j$ would act horizontally. 

\begin{figure}[hbtp]
\centering
\includegraphics[scale=1]{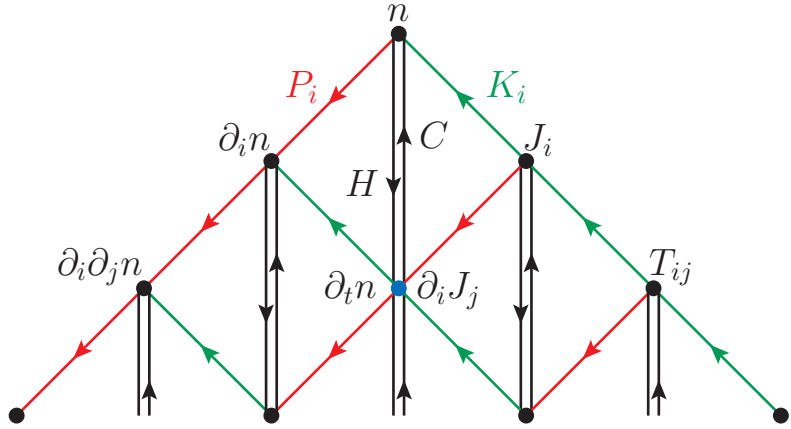}
\caption{The family tree of the operator $n$. }
\label{fig:n_pyramid}
\end{figure}

As an example, we consider the particle number operator $n=\psi^\dagger\psi$, a primary with particle number zero. Using the notation $ \overleftrightarrow{\d_i}=\overrightarrow{\d_i}-\overleftarrow{\d_i}$, the first two alien operators are $J_i=-\frac{i}{2} \psi^\+\overleftrightarrow{\d_i}\psi$ and  $T_{ij}=\frac{-1}{4} \psi^\+\overleftrightarrow{\d_i}\overleftrightarrow{\d_j}\psi$, which appear on the far right at each level\footnote{We note that the operator $\psi^\dagger \overleftrightarrow{\d_t} \psi$ is also an alien operator which lowers in the same way as $\delta^{ij}T_{ij}$. However, since there is a linear combination of the two which is a primary, we need only include one of the two in the family of the operator $n$ and of course we pick $T_{ij}$ since it carries more information than just its trace. A similar story happens at every level of the tree.}. Successive applications of $K_i$ and $P_i$ take us up or down along the tree but always slanted to the left (see fig. \ref{fig:n_pyramid}). We note that by a simple counting argument, all possible spatial derivative combinations of $\psi$ and $\psi^\+$ appear on the tree.

In the family tree of a generic operator with particle number zero, each dot on the diagram only the position of an operator with respect to other operators. A remark is in order for when a spot on the diagram is occupied by more than one operator, which is the case for most positions. Two scenarios can arise. First, it is possible that there is no further relation between these operators than what is derived from their position on the family tree. That is, they are in fact distinct operators. This is the case for the descendants of a generic neutral operator.

However, it is also possible that there is a linear combination of the operators sitting on the same spot that lowers to zero using both $C$ and $K_i$. This is similar to the case of the null operator in the previous. As an example, again we specialize to the family of the operator $n$ (However, the same arguments can be applied in any family where these requirements are met). Consider the spot denoted by the blue dot in figure \ref{fig:n_pyramid}. Here we have the two operators $\d_t n$ and $\d_i J_j$ which are of course related by the continuity equation $\d_tn+\d_i J_i=0$. And it is in fact this linear combination that is null\footnote{However, because of the existence of the alien operator $T_{ij}$ this null combination does not restrict the form of the OPE as in the previous section.}. 

For the case of the primary operator $n$, it turns out that there is a single alien operator at each level of the family tree (e.g. $J_i$ and $T_{ij}$ for levels one and two) and that there exists a null operator (in these cases $\d_t n + \d_i J_i$ and $\d_t J_j + \d_i T_{ij}$ respectively.) Here, we give the explicit form of the alien operator entering at level $l$:
\begin{equation}
	C^{(l)}_{i_1\cdots i_l}= (2i)^{-l}
\psi^\+\overleftrightarrow{\d_{i_1}}\cdots \overleftrightarrow{\d_{i_l}} \psi
\end{equation}
We define $C^{(0)}\equiv n$ and note that $C^{(1)}_i=J_i$ and $C^{(2)}_{ij}=T_{ij}$. Using this definition, we have:
\begin{equation}
[K_k,C^{(l)}_{i_1\cdots i_l}]=
	i\sum\limits_{n=1}^l \delta_{kn}C^{(l-1)}_{i_1\cdots \overline{i_n}\cdots i_l}
	\;\;,\;\;\;\;\; [C,C^{(l)}_{i_1\cdots i_l}]=0,
\end{equation}
where the notation $\overline{i_n}$ implies the $n$'th index is ommited. With these definitions for the alien operator at level $l$, we can easily derive that the following combination is null:
\begin{equation}
	\label{eq:null_ops}
	O^{(l)}=\d_t C^{(l)}_{i_1\cdots i_l}+\d_j C^{(l+1)}_{j i_1\cdots i_l}.
\end{equation}

 The fact that this null combination  has  the form of a conservation law is very suggestive. Whether or not the operator $O^{(l)}$ is in fact zero is another matter which cannot be answered by looking at the operator algebra alone. In general we can show that $O^{(1)}=O^{(2)}=0$, which are nothing but the continuity and energy conservation equations. In the free theory one can also demonstrate that the infinite conservation laws in this sector in fact arise from the $O^{(l)}$'s \cite{moroz:11}. It is also known that in the interacting theory all but the first few are broken.

What the algebra does demonstrate is that if there are conservation laws, they should be found among the null operators which relate the alien operators of adjacent levels. And in fact if there is a conservation law in the free theory that is broken in the interacting theory, its non-conservation must act as a primary operator.  That is, its n-point functions with other primaries are restricted as in equations \eqref{eq:2pt} and \eqref{eq:3-4pt}.

\section{Conclusion}
We have shown that it is possible to further restrict the OPE and n-point functions of some primary operators in CFTs just by algebra considerations. In particular, we showed that operators with critical scaling dimension $d/2$, e.g. the elementary particle $\psi$, has known OPE with any other primary. Because of this, its 3-point function is determined up to an overall constant.

We also analyzed the structure of the descendants of primaries with zero particle number and showed that there are non-descendant operators that nevertheless play an important role in the family, the so-called alien operators. There is an intimate relationship between null operators derived from these alien operators and conservation laws. However, the question of conservation is one that needs to be looked at in each theory.

\subsection*{Acknowledgments}
This work is supported, in part, by DOE grant DE-FG02-13ER41958. The authors thank Sergej Moroz for extensive discussion and comments.

\bibliographystyle{utphys}
\bibliography{biblio}

\providecommand{\href}[2]{#2}\begingroup\raggedright\begin{thebibliography}{10}

\bibitem{Hagen:72}
C.~R. Hagen, \href{http://dx.doi.org/10.1103/PhysRevD.5.377}{``Scale and
  conformal transformations in galilean-covariant field theory,''{\em Phys.
  Rev. D} {\bf 5} (Jan, 1972)  377--388}.
  \url{http://link.aps.org/doi/10.1103/PhysRevD.5.377}.

\bibitem{Niederer:72}
U.~Niederer, ``{The maximal kinematical invariance group of the free
  Schrodinger equation.},''
{\em Helv.Phys.Acta} {\bf 45} (1972)  802--810.

\bibitem{Mehen:99}
T.~Mehen, I.~W. Stewart, and M.~B. Wise, ``{Conformal invariance for
  nonrelativistic field theory},''
  \href{http://dx.doi.org/10.1016/S0370-2693(00)00006-X}{{\em Phys.Lett.} {\bf
  B474} (2000)  145--152},
\href{http://arxiv.org/abs/hep-th/9910025}{{\tt arXiv:hep-th/9910025
  [hep-th]}}.

\bibitem{Henkel:93}
M.~Henkel, ``{Schrodinger invariance in strongly anisotropic critical
  systems},'' \href{http://dx.doi.org/10.1007/BF02186756}{{\em J.Statist.Phys.}
  {\bf 75} (1994)  1023--1061},
\href{http://arxiv.org/abs/hep-th/9310081}{{\tt arXiv:hep-th/9310081
  [hep-th]}}.

\bibitem{Henkel:03}
M.~Henkel and J.~Unterberger, ``{Schrodinger invariance and space-time
  symmetries},'' \href{http://dx.doi.org/10.1016/S0550-3213(03)00252-9}{{\em
  Nucl.Phys.} {\bf B660} (2003)  407--435},
\href{http://arxiv.org/abs/hep-th/0302187}{{\tt arXiv:hep-th/0302187
  [hep-th]}}.

\bibitem{Son:07}
Y.~Nishida and D.~T. Son, ``{Nonrelativistic conformal field theories},''
  \href{http://dx.doi.org/10.1103/PhysRevD.76.086004}{{\em Phys.Rev.} {\bf D76}
  (2007)  086004},
\href{http://arxiv.org/abs/0706.3746}{{\tt arXiv:0706.3746 [hep-th]}}.

\bibitem{Son:08}
D.~Son, ``{Toward an AdS/cold atoms correspondence: A Geometric realization of
  the Schrodinger symmetry},''
  \href{http://dx.doi.org/10.1103/PhysRevD.78.046003}{{\em Phys.Rev.} {\bf D78}
  (2008)  046003},
\href{http://arxiv.org/abs/0804.3972}{{\tt arXiv:0804.3972 [hep-th]}}.

\bibitem{McGreevy:08}
K.~Balasubramanian and J.~McGreevy, ``{Gravity duals for non-relativistic
  CFTs},'' \href{http://dx.doi.org/10.1103/PhysRevLett.101.061601}{{\em
  Phys.Rev.Lett.} {\bf 101} (2008)  061601},
\href{http://arxiv.org/abs/0804.4053}{{\tt arXiv:0804.4053 [hep-th]}}.

\bibitem{Moroz:09}
C.~A. Fuertes and S.~Moroz, ``{Correlation functions in the non-relativistic
  AdS/CFT correspondence},''
  \href{http://dx.doi.org/10.1103/PhysRevD.79.106004}{{\em Phys.Rev.} {\bf D79}
  (2009)  106004},
\href{http://arxiv.org/abs/0903.1844}{{\tt arXiv:0903.1844 [hep-th]}}.

\bibitem{Volovich:09}
A.~Volovich and C.~Wen, ``{Correlation Functions in Non-Relativistic
  Holography},'' \href{http://dx.doi.org/10.1088/1126-6708/2009/05/087}{{\em
  JHEP} {\bf 0905} (2009)  087},
\href{http://arxiv.org/abs/0903.2455}{{\tt arXiv:0903.2455 [hep-th]}}.

\bibitem{Zimmermann:70}
W.~Zimmermann, ``Local operator products and renormalization in quantum field
  theory,'' {\em Lectures on Elementary Particles and Quantum Field Theory,}
  (1970)  .

\bibitem{Wilson:71}
K.~G. Wilson, ``{Renormalization group and critical phenomena. 2. Phase space
  cell analysis of critical behavior},''
\href{http://dx.doi.org/10.1103/PhysRevB.4.3184}{{\em Phys.Rev.} {\bf B4}
  (1971)  3184--3205}.

\bibitem{Wilson:72}
K.~G. Wilson and W.~Zimmermann, ``Operator product expansions and composite
  field operators in the general framework of quantum field theory,'' {\em
  Communications in Mathematical Physics} {\bf 24} (1972) no.~2, 87--106.
  \url{http://projecteuclid.org/euclid.cmp/1103857739}.

\bibitem{Dobrev:97}
V.~Dobrev, H.-D. Doebner, and C.~Mrugalla, ``Lowest weight representations of
  the schrödinger algebra and generalized heat/schrödinger equations,''
  \href{http://dx.doi.org/http://dx.doi.org/10.1016/S0034-4877(97)88001-9}{{\em
  Reports on Mathematical Physics} {\bf 39} (1997) no.~2, 201 -- 218}.
  \url{http://www.sciencedirect.com/science/article/pii/S0034487797880019}.

\bibitem{Dobrev:13}
V.~Dobrev, ``{Non-Relativistic Holography - A Group-Theoretical Perspective},''
  \href{http://dx.doi.org/10.1142/S0217751X14300014}{{\em Int.J.Mod.Phys.} {\bf
  A29} (2014)  1430001},
\href{http://arxiv.org/abs/1312.0219}{{\tt arXiv:1312.0219 [hep-th]}}.

\bibitem{moroz:11}
X.~Bekaert, E.~Meunier, and S.~Moroz, ``{Symmetries and currents of the ideal
  and unitary Fermi gases},''
  \href{http://dx.doi.org/10.1007/JHEP02(2012)113}{{\em JHEP} {\bf 1202} (2012)
   113},
\href{http://arxiv.org/abs/1111.3656}{{\tt arXiv:1111.3656 [hep-th]}}.

\end{thebibliography}\endgroup

\end{document}